\documentclass[pra,twocolumn,showpacs,superscriptaddress,aps]{revtex4-1}
\usepackage{amsfonts}
\usepackage{amsmath}
\usepackage{amssymb}
\usepackage{graphicx}
\usepackage{dcolumn}
\usepackage{times}

\begin{document}

\title{Matter-wave solitons in the counterflow of two immiscible superfluids}
\author{F. Tsitoura}
\affiliation{Department of Physics, University of Athens,
Panepistimiopolis, Zografos, Athens 15784, Greece}
\author{V. Achilleos}
\affiliation{Department of Physics, University of Athens,
Panepistimiopolis, Zografos, Athens 15784, Greece}
\author{B. A. Malomed}
\affiliation{Department of Physical Electronics, School of
Electrical Engineering, Faculty of Engineering, Tel Aviv University,
Tel Aviv 69978, Israel}
\author{D. Yan}
\affiliation{Department of Mathematics and Statistics, University of
Massachusetts, Amherst MA 01003-4515, USA}
\author{P. G. Kevrekidis}
\affiliation{Department of Mathematics and Statistics, University of
Massachusetts, Amherst MA 01003-4515, USA}
\author{D. J. Frantzeskakis}
\affiliation{Department of Physics, University of Athens,
Panepistimiopolis, Zografos, Athens 15784, Greece}

\begin{abstract}

We study formation of solitons induced by counterflows of immiscible
superfluids. Our setting is based on a quasi-one-dimensional binary
Bose-Einstein condensate (BEC), composed of two immiscible
components with large and small numbers of atoms in them. Assuming
that the ``small" component moves with constant velocity, either by
itself, or being dragged by a moving trap, and intrudes into the
``large" counterpart, the following results are obtained. Depending
on the velocity, and on whether the small component moves in the absence or in
the presence of the trap, two-component dark-bright solitons, scalar
dark solitons, or multiple dark solitons may emerge, the latter
outcome taking place due to breakdown of the superfluidity. We
present two sets of analytical results to describe this
phenomenology. In an intermediate velocity regime, where dark-bright
solitons form, a reduction of the two-component Gross-Pitaevskii
system to an integrable Mel'nikov system is developed, demonstrating
that solitary waves of the former are very accurately described by
analytically available solitons of the latter. In the high-velocity
regime, where the breakdown of the superfluidity induces the
formation of dark solitons and multi-soliton trains, an effective
single-component description, in which a strongly localized wave
packet of the ``small" component acts as an effective potential for
the ``large" one, allows us to estimate the critical velocity beyond
which the coherent structures emerge in good agreement with the
numerical results.

\end{abstract}

\maketitle

\section{Introduction}

Matter-wave solitons is a theme that has attracted much attention in studies
of atomic Bose-Einstein condensates (BECs) \cite{book2a,book2} and, more
generally, in studies of superfluids \cite{volo} -- see, e.g., recent
reviews \cite{emergent,revnonlin,djf}. In atomic BECs, matter-wave solitons
have been realized by means of various methods, including phase-imprinting
\cite{dark1}, density engineering \cite{harv}, quantum-state engineering
\cite{hamburg}, transport in optical lattices \cite{scot1}, transport past
single defects \cite{engelsath} or through disordered potentials \cite%
{disord}, and nonlinear interference between individual condensates \cite%
{collisions}. Furthermore, matter-wave solitons have also been observed in
multi-component BECs consisting of two different spin states of the same
atom species. In particular, atomic dark-bright solitons have been realized
in binary rubidium condensates by means of the quantum-state engineering
technique \cite{hamburg}, or the counterflow between two miscible components
\cite{engtrain1,engtrain2}; employing the latter technique, generation of dark-dark
solitons was reported too \cite{darkdark1,darkdark2}. Importantly, solitons have been
argued to emerge spontaneously upon crossing the BEC phase transition, via
the Kibble-Zurek mechanism \cite{zurek}. This can be thought of as a
quasi-one-dimensional analog of the pioneering experiments on the same
mechanism in higher-dimensional BECs performed in~\cite{BPA}.

As mentioned above, the counterflow of two miscible,
quasi-one-dimensional (1D) BEC superfluids, is one of the techniques
that were recently used for the generation of matter-wave solitons.
More generally, the counterflow dynamics of superfluids is an
interesting subject due to its links with such fundamental phenomena
as turbulence, pattern formation, and generation of topological
defects, which occur in a vast variety of physical contexts
ranging from superfluid helium to cosmological strings and inflation 
\cite{defects}. In the context of binary atomic BECs, a number of theoretical
works have been dealing with the counterflow of either miscible 
\cite{tsubota1,tsubota2} or immiscible \cite{marklund1,marklund2} binary
superfluids. In the former case, quantum turbulence and vortex generation in
higher-dimensional settings \cite{tsubota1} and in spinor BECs with $F=1$
\cite{tsubota2} were studied; in the latter case, effects of quantum
swapping, Rayleigh-Taylor instability \cite{marklund1}, and parametric
resonance of capillary waves were predicted \cite{marklund2}. Furthermore, 
more recently, the counterflow superfluid of polaron pairs in 
Bose-Fermi mixtures in optical lattices was analyzed \cite{BF}.

In the present work, we propose and analyze a physically relevant
setting based on the counterflow of two immiscible superfluids. This
setting can be used for the generation of scalar or vector solitons
(of the dark-bright type) in binary BECs, and also for studies of
their superfluid properties. We consider a quasi-1D binary BEC
mixture composed of two immiscible components, ``small" and ``large"
ones, with ratios of axial sizes and number of atoms in them $\sim
1:100$. We then assume that the ``small" component moves with
constant velocity $\upsilon $ through the ``large" one, by itself or
being dragged by a moving trapping potential. 

Systematic simulations of this setting produce the following phenomenology.
For sufficiently small dragging velocities, $\upsilon <0.61\tilde{c}_{s}$,
with $c_{s}\equiv \sqrt{2}\tilde{c}_{s}$ being the speed of sound of the
large component, and if the trap of the small component is switched off, the
small component cannot remain localized, hence no well-defined patterns are
formed. For intermediate velocities, $0.61\tilde{c}_{s}<\upsilon <0.95\tilde{%
c}_{s}$, a robust two-component dark-bright soliton is formed. It is
composed of a bright (dark) soliton in the small (large) component
and can be very well approximated analytically. In particular, we
use a multiscale asymptotic expansion method to reduce the
underlying system of the Gross-Pitaevskii equations (GPEs) to a
completely integrable Mel'nikov system \cite{Mel}, thus very
accurately mapping the dark-bright solitary wave, which
spontaneously emerges in the GPE system, into an exact soliton of
the Mel'nikov system. On the other hand, for
$0.58\tilde{c}_{s}<\upsilon <0.87\tilde{c}_{s}$ but in the presence
of its trap dragging the small component, the ``Mel'nikov's"
dark-bright soliton still
emerges but is unstable and eventually decays. For larger trap velocities, $%
\upsilon >0.95\tilde{c}_{s}$, and if the trap is absent, a scalar dark
soliton is created in the large component. Furthermore, for $\upsilon >0.87%
\tilde{c}_{s}$ and in the presence of the trap, the small component is
deformed into a sharply localized object, with a width on the order of the
healing length, which propagates with a velocity larger than the critical
velocity for the breakdown of the superfluidity in the large component. The
breakdown leads to the emergence of multiple dark solitons, similarly to the
results reported in previous theoretical \cite{hakim,thsf} and experimental
\cite{engelsath} works (cf. also the quasi-two-dimensional analog of such an
experiment producing vortex dipoles~\cite{BPA2}).

The rest of the paper is structured as follows. The model is formulated in
Section II. Then, in Section III, we systematically analyze, by means of
analytical approximations and direct simulations, two different scenarios of
the evolution of the small component, in the absence or in the presence of
its trap. Finally, in Section IV we summarize our findings and discuss
possibilities for the extension of the analysis.

\section{The model and setup}

We consider the binary BEC composed of two hyperfine states of $^{87}$Rb,
that are described by macroscopic wave functions $\Psi _{j}(\mathbf{r},t)$ ($%
j=1,2$). The dynamics can then be described by the following system of
coupled GPEs \cite{book2a,book2}:
\begin{equation}
i\hbar \partial _{t}\Psi _{j}=\left( -\frac{\hbar ^{2}}{2m}%
\nabla ^{2}+V_{j}(\mathbf{r})-\mu _{j}+\sum_{k=1}^{2}g_{jk}|\Psi
_{k}|^{2}\right) \Psi _{j},  
\label{oldeq}
\end{equation}
where $m$ is the atomic mass, $\mu _{j}$ are chemical potentials, $V_{j}(%
\mathbf{r})$ are trapping potentials, and $g_{jk}=4\pi \hbar ^{2}a_{jk}/m$
are coupling constants defined by the $s$-wave scattering lengths, $a_{jk}$.
Notice that, for states $|1,-1\rangle $ and $|2,1\rangle $, which were used
in the experiment of Ref.~\cite{mertes}, or states $|1,-1\rangle $ and $%
|2,-2\rangle $ used in Refs.~\cite{engtrain1,engtrain2}, all the scattering lengths
take approximately equal (positive) values. Below we assume $%
a_{11}=a_{22}\neq a_{12}$, hence the sign of parameter $\Delta \equiv
(a_{11}a_{22}-a_{12}^{2})/a_{11}^{2}$, which determines whether the two
components are miscible ($\Delta >0$) or immiscible ($\Delta <0$) \cite%
{mineev}, depends only on ratio $\beta \equiv a_{12}/a_{11}$: if $\beta >1$ (%
$\beta <1$), then the two superfluids are immiscible (miscible).

We consider the case when both components are confined in strongly
anisotropic (quasi-1D) traps, which have the form of rectangular boxes of
lengths $L_{x_{j}}\gg L_{y_{j}}=L_{z_{j}}\equiv L_{\perp }$, with the
transverse-confinement lengths $L_{\perp _{j}}$ being on the order of the
healing lengths $\xi _{j}$. For such a highly anisotropic trap, it is
relevant to factorize the wave functions into longitudinal and transverse
components, namely $\{u(x,t),\mathrm{v}(x,t)\}$ and $\Phi _{j}(y,z)$ (see,
e.g., Refs.~\cite{emergent,revnonlin,djf}): 
\begin{eqnarray}
\Psi _{1}(\mathbf{r},t) &=&\Phi _{1}(y,z)u(x,t)\exp (-iE_{1}t/\hbar ),
\label{decomp1} \\
\Psi _{2}(\mathbf{r},t) &=&\Phi _{2}(y,z)\mathrm{v}(x,t)\exp (-iE_{2}t/\hbar
),  \label{decomp2}
\end{eqnarray}
where the transverse modal functions $\Phi _{j}(y,z)$ and energies $E_{j}$
are determined by the auxiliary 2D problem for the quantum oscillator in the
transverse box. Since the considered system is effectively one-dimensional,
it is natural to assume that $\Phi _{j}$ remains in the ground state, i.e., $%
\Phi _{j}(y,z)=(\sqrt{2}/L_{\perp })\mathrm{sin}\left( \pi y/L_{\perp
}\right) \mathrm{sin}\left( \pi z/L_{\perp }\right) $, for $j=1,2$. Using
these expressions, we substitute expressions~(\ref{decomp1})-(\ref{decomp2})
into Eqs.~(\ref{oldeq}) and perform averaging over the transverse
coordinates, to derive a system of scaled equations for the longitudinal
wave functions $u(x,t)$ and $\mathrm{v}(x,t)$ (see also Ref.~\cite{ccr}):
\begin{eqnarray}
iu_{t} &=&-u_{xx}+(|u|^{2}+\beta |\mathrm{v}|^{2}-\mu _{1})u+V_{1}(x)u,
\label{oldeq1} \\
i\mathrm{v}_{t} &=&-\mathrm{v}_{xx}+(\beta |u|^{2}+|\mathrm{v}|^{2}-\mu _{2})%
\mathrm{v}+V_{2}(x)\mathrm{v},  \label{oldeq2}
\end{eqnarray}
where subscripts stand for partial derivatives. In these equations,
longitudinal coordinate $x$, time $t$, densities $|u|^{2}$, 
$|\mathrm{v}|^{2}$, and energy are measured, respectively, 
in units of the healing length 
$\xi _{1}\equiv \hbar /\sqrt{2mn_{1}\tilde{g}_{11}}$, 
characteristic time $\sqrt{2}\xi _{1}/c_{s}$ 
(where $c_{s}\equiv \sqrt{\tilde{g}_{11}n_{1}/m}$ is the sound speed 
of the first component), density $n_{1}$, and characteristic energy 
$\tilde{g}_{11}n_{1}$ of the first component. 
Finally, $\tilde{g}_{ij}=(9/4L_{\perp }^{2})g_{ij}$ are the effective 1D
interaction strengths.

\begin{figure}[tbp]
\centering
\includegraphics[scale=0.37]{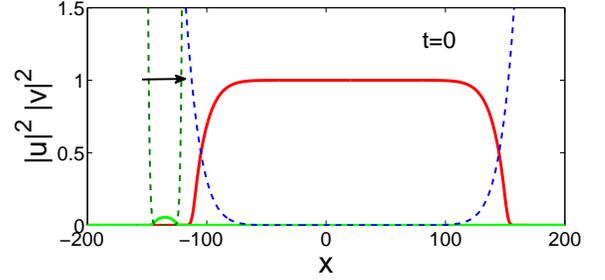}
\caption{(Color online) The density profiles (solid lines) of the two
immiscible superfluids, confined by the box-shaped traps $V_{1}(x)$ and 
$V_{2}(x,t)$ (dashed lines) at $t=0$. The red (green) line corresponds to the
$u$- ($\mathrm{v}$-) component. The arrow depicts the direction of motion of
the $\mathrm{v}$-component at $t>0$.}
\label{fig1}
\end{figure}

It is now useful to adopt experimentally relevant values of the parameters.
First, we fix the scattering length and the density of first component to 
$a_{11}=5.77$~nm and $n_{1}=10^{8}\mathrm{m}^{-1}$, which results in the
healing length, $\xi _{1}=0.25~\mu $m; next, we fix $\beta =1.1$, which
corresponds to the immiscibility (we have checked that other values of 
$\beta >1$ lead to qualitatively similar results). We will assume that the
first component ($u$) is ``much larger" than the other one 
($\mathrm{v}$), in the sense that the respective ratio of the numbers of atoms
is $N_{1}/N_{2}\approx 100$; accordingly, the chemical potentials are fixed
to $\mu _{1}=1$ and $\mu _{2}=0.06$. As concerns the size of the traps, we
assume that, in most cases, $L_{x_{1}}=57.6~\mu $m, $L_{x_{2}}=5~\mu $m, and
$L_{\perp }=0.8~\mu $m, which correspond, respectively, to $L_{x_{1}}=244$, 
$L_{x_{2}}=21$ and $L_{\perp }=3$ in the dimensionless units; an exception
concerns Fig.~\ref{fig5} (see below), where we use $L_{x_{1}}=123~\mu $m (or
$490$ in the dimensionless units) for the length of the $u$-component. We
have checked that, as long as the ratios of the atom numbers, chemical
potentials and trap lengths are such that 
$N_{1}/N_{2}\sim \mu _{1}/\mu_{2}\sim 100$ 
and $L_{x_{1}}:L_{x_{2}}:L_{\perp }\sim 100:10:1$, 
the results are qualitatively similar.

In our simulations, we will approximate the box-like trapping potentials 
$V_{j}$ by super-Gaussians:
\begin{equation}
V_{j}=V_{0}\left[ 1-\mathrm{\exp }\left( -\left( \frac{x+x_{j}}{w_{j}}%
\right) ^{12}\right) \right] ,
\end{equation}
where $V_{0}$ is the (common) trap amplitude for both components, while 
$x_{j}$ and $w_{j}$ denote the positions and widths of the traps,
respectively. We fix these parameters as $V_{0}=10$, $x_{1}=-20$, 
$w_{1}=160$, and $w_{2}=20$; as concerns the position of the trap acting 
on the $\mathrm{v}$-component, it is assumed to be initially (at $t=0$) 
placed at $x_{2}(0)=135$ and move at a constant velocity: 
$x_{2}=135-\upsilon t$. Thus, the ``small" $\mathrm{v}$-component 
starts moving to the right, penetrating the ``large" $u$-component; 
notice that $\upsilon$ is measured in units of 
$\tilde{c}_{s}=c_{s}/\sqrt{2}$ (recall that $c_{s}$ is the speed of 
sound of the $u$-component). As we show
below, such an effective counterflow gives rise to the formation of
nonlinear structures and solitons that we will investigate in detail
in the following section. Furthermore, we note in passing that in the 
immiscible case considered here, the counterflow
with speed $\upsilon$ is modulationally stable, 
in accordance with the criterion of \cite{engtrain1,darkdark1}.
This can also be intuitively understood on the basis of immiscibility 
of two independently modulationally stable components.

\begin{figure}[tbp]
\centering
\includegraphics[scale=0.37]{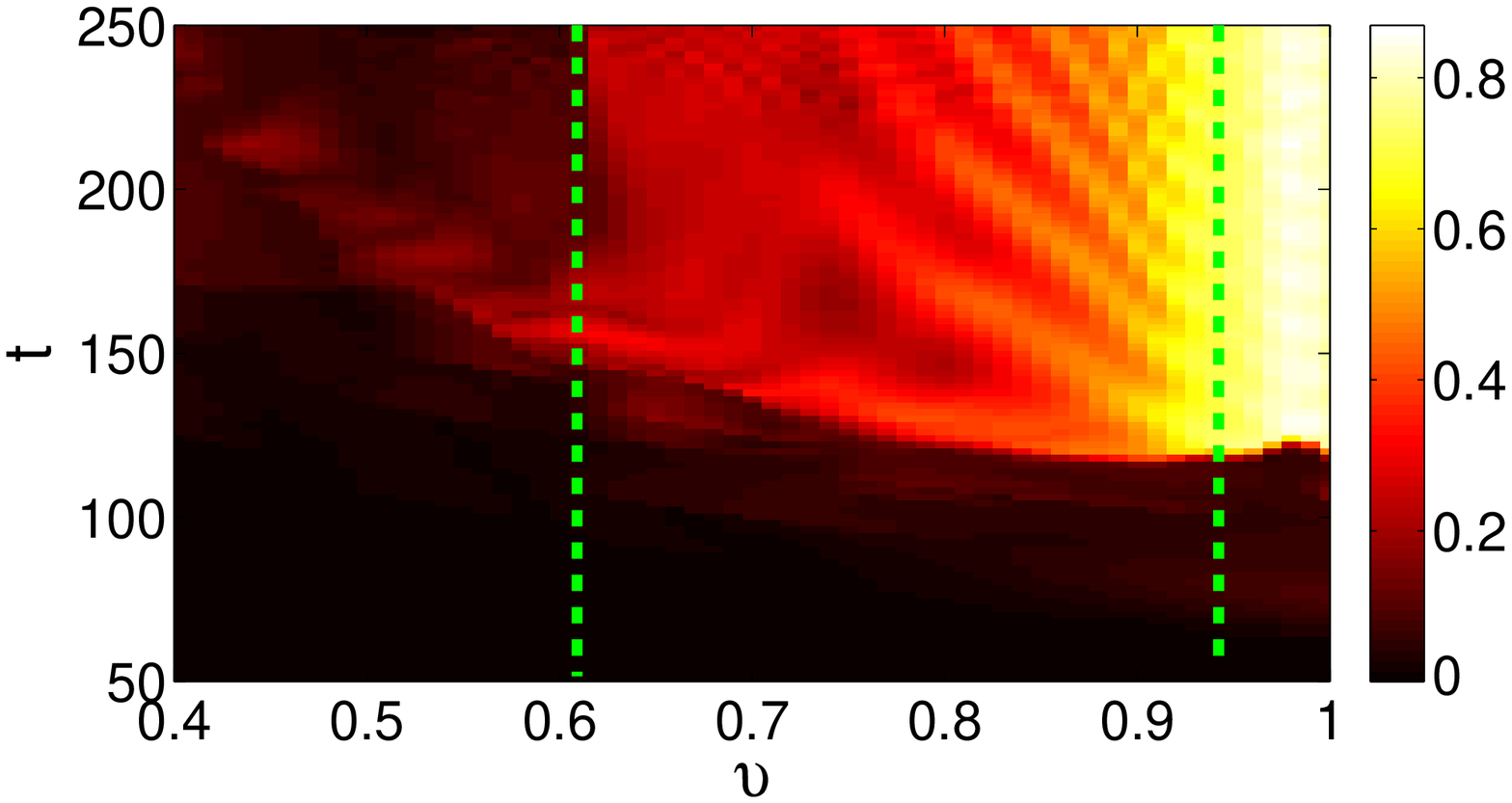} %
\includegraphics[scale=0.37]{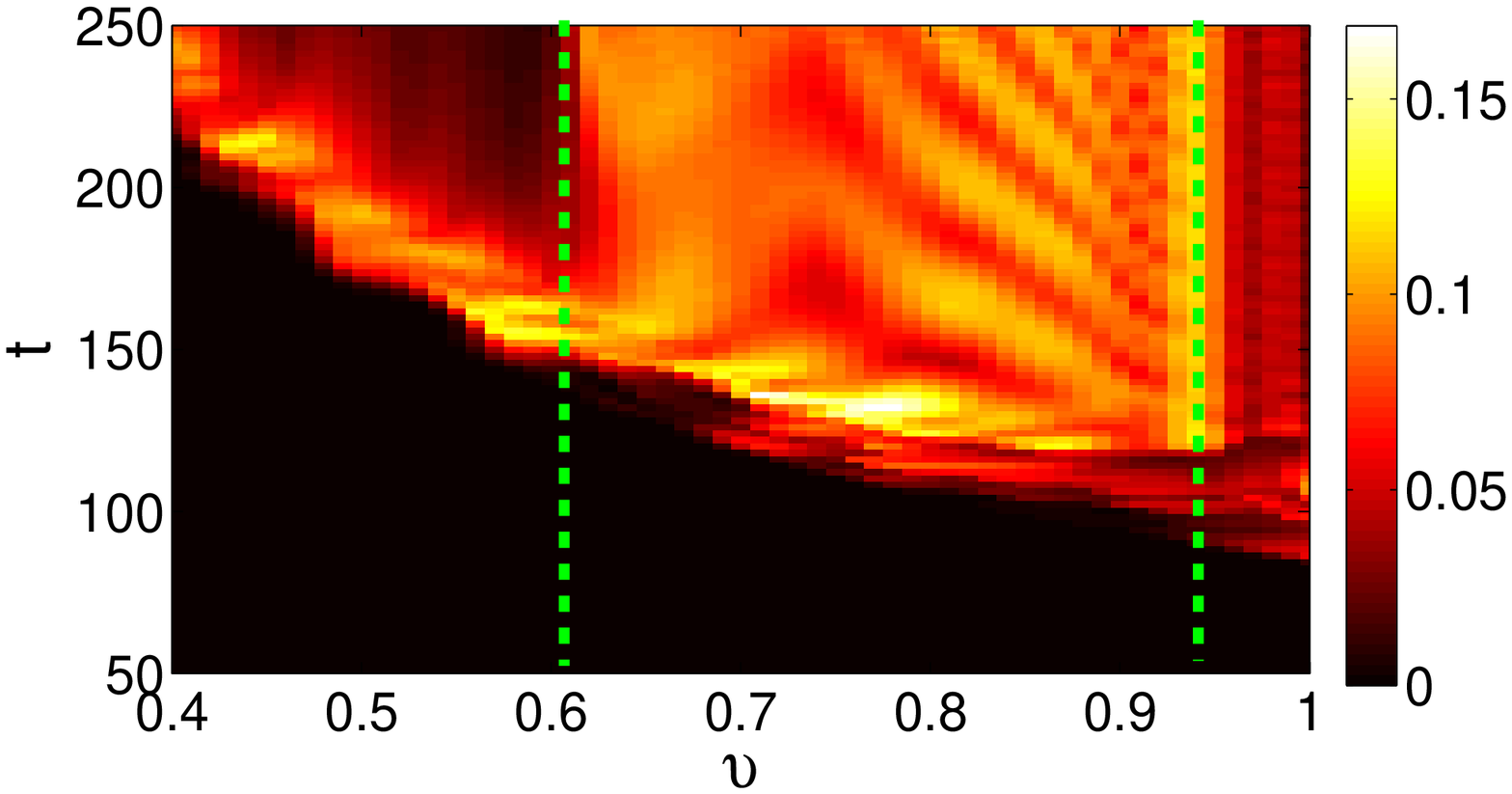}
\caption{(Color online) Contour plots showing the evolution of the largest
density dip in the ${u}$-component (the top panel) and the peak density of
the $\mathrm{v}$-component (the bottom panel), with respect to the velocity
of the motion of the trap. We identify three different regimes, separated by
the vertical (green) dashed lines: a) $0<\protect\upsilon <0.61$, 
where no localized excitations are formed, b) 
$0.61<\protect\upsilon <0.95$, where traveling dark-bright solitons
emerge, and c) $\protect\upsilon >0.95$, where deep
dark solitons are
formed in the ${u}$-component. In the latter case, the $\mathrm{v}$%
-component is quickly dispersed.}
\label{fig2}
\end{figure}

The initial configuration, obtained by means of an imaginary-time
integration, is depicted in Fig.~\ref{fig1}, where the two components are
initially separated. Then, at $t>0$, the $\mathrm{v}$-component is set into
motion and moves through the $u$-component. The evolution of the system is
monitored by numerically integrating Eqs.~(\ref{oldeq1}) and (\ref{oldeq2})
in real time by means of the split-step Fourier method.

\section{Analytical and numerical results}

We will consider two different cases: (A) when trap $V_{2}$, acting on the
small component, is switched off after it has penetrated the $u$-component
(for the above-mentioned parameters, this occurs at $x=-40$ and 
$t\sim 100$); (B) the trap $V_{2}$ is not switched off, i.e., the 
$\mathrm{v}$-component is permanently dragged by its trap.

\subsection{Evolution of the small  
component in the absence of the trap.}

In this case, our systematic simulations have revealed the existence of
three different velocity regimes, characterized by the emergence (or the
failure to emerge) of various localized nonlinear structures. These regimes
were identified by calculating -- at each time step -- the maximum density
dip in the ${u}$-component, as well as the peak density of the $\mathrm{v}$%
-component, as shown in the top and bottom panel of Fig.~\ref{fig2},
respectively.

\begin{figure}[tbp]
\centering
\includegraphics[scale=0.37]{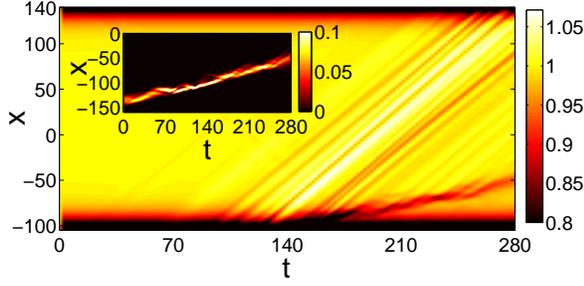}
\caption{(Color online) Contour plots showing the evolution of the
density of the $u$-component (the inset shows
the evolution of the $\mathrm{v}$-component) for trap's velocity 
$\protect \upsilon =0.3$. 
It is observed, aside from the emission
of sound waves in the $u$-component, that a structure resembling a
dark-bright solitary wave is formed. However, it slowly disperses in
the course of the evolution.} \label{fig3}
\end{figure}

In particular, for velocities $0<\upsilon <0.61$ 
[corresponding to the region on the left of the first dashed (green) 
line in Fig.~\ref{fig2}],
no clear solitary-wave structure is formed in the dynamics. There exist,
however, numerous small-amplitude dips that are formed in the 
${u}$-component, which propagate with speeds approximately equal to the speed of
sound in that component. Generally, for such relatively small velocities, as
long as the $\mathrm{v}$-component penetrates the $u$-component, strong
emission of radiation is observed. A typical example is shown in 
Fig.~\ref{fig3}, for $\upsilon =0.3$. In addition to the small-amplitude
sound wavepackets propagating in the $u$ component, there exists a
slow structure (with a speed near close to that of the initially
moving trap), which emerges with a larger dip in the $u$-component
and a bright ``bump" in the $v$ component. Yet, this structure does
not preserve its shape, but rather slowly disperses, contrary
to what would be expected of a robust solitary wave. 

\begin{figure}[tbp]
\centering
\includegraphics[scale=0.37]{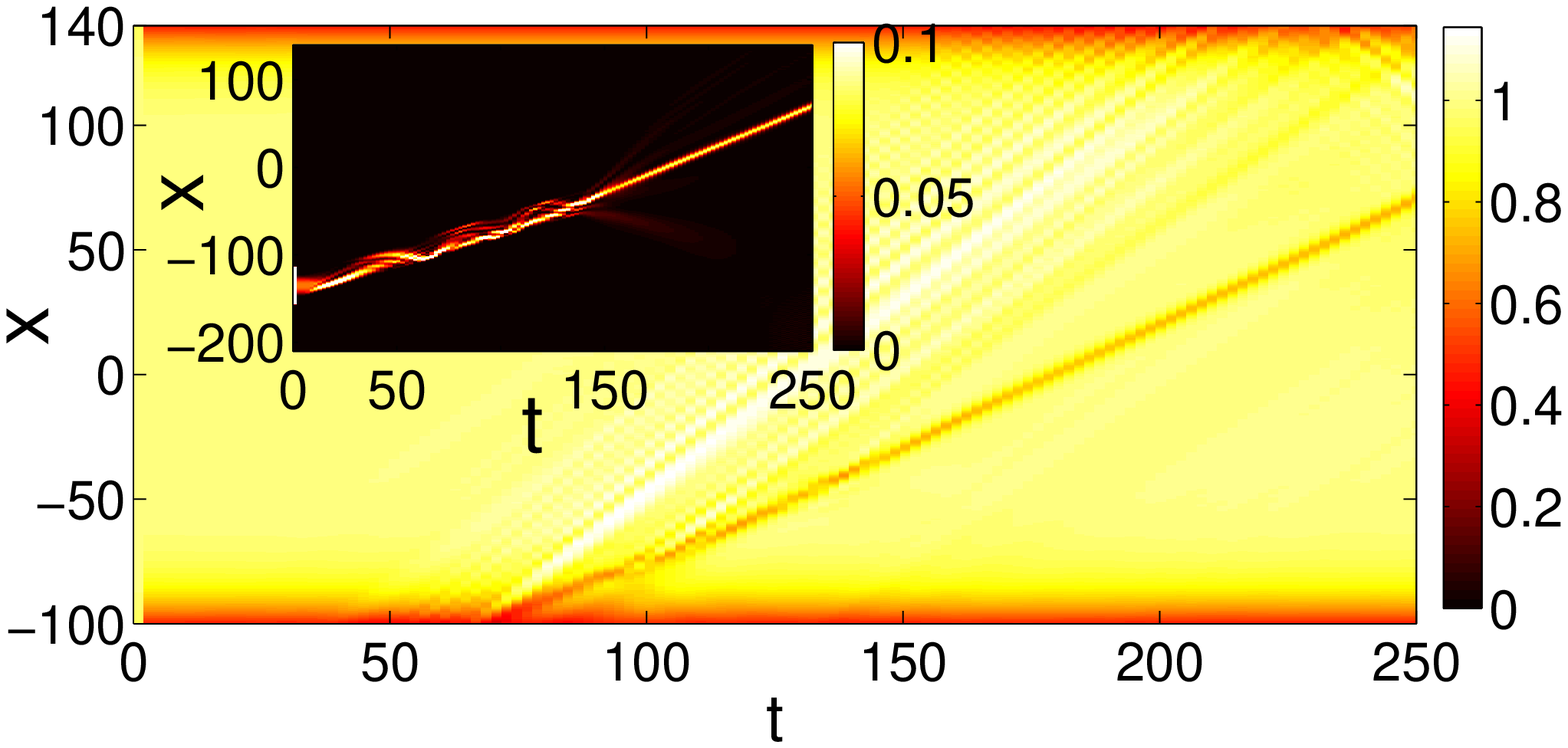} %
\includegraphics[scale=0.37]{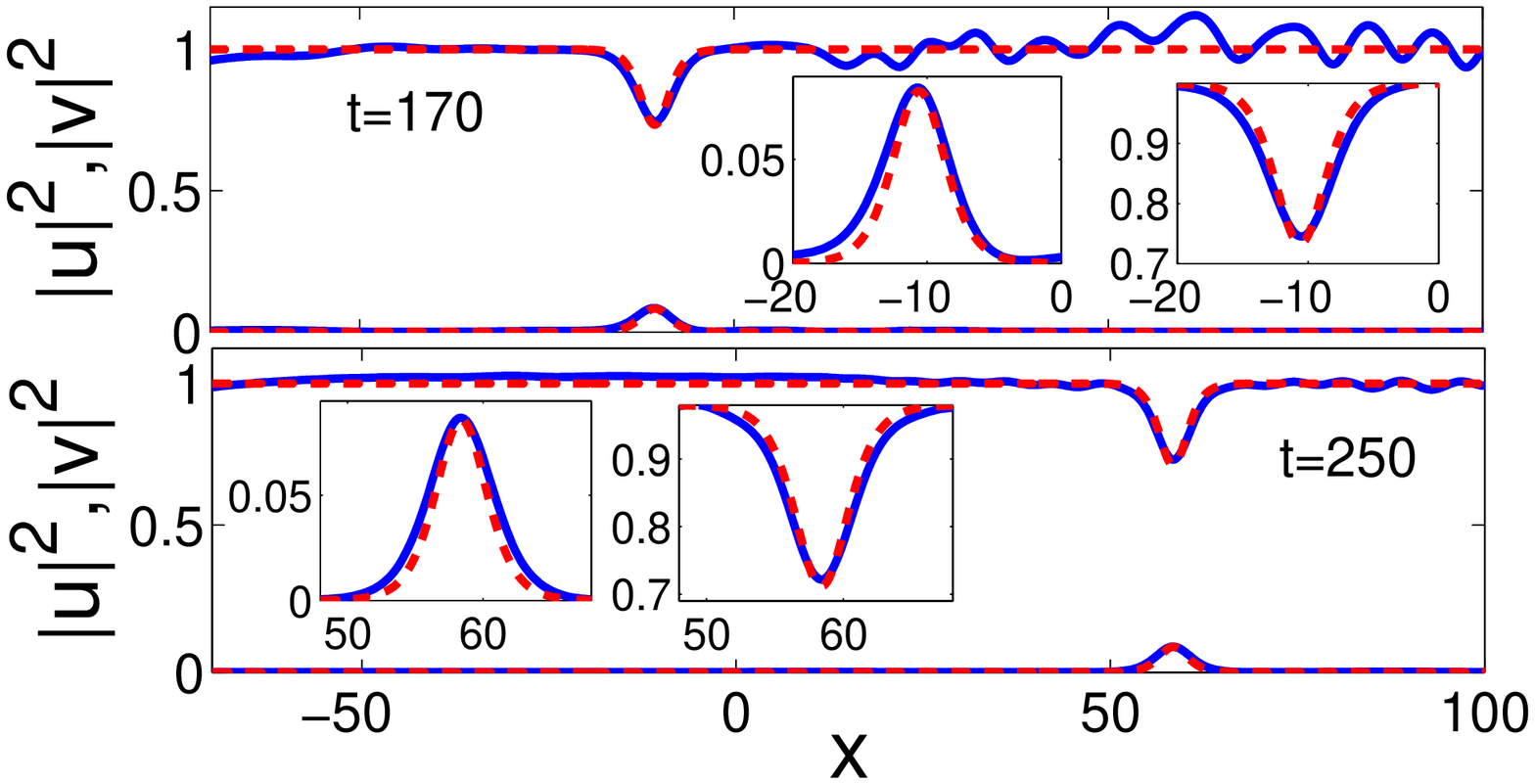}
\caption{(Color online) The top panel shows the evolution of the density the
$u$-component (inset shows the evolution of the $\mathrm{v}$-component) for
trap's velocity $\protect\upsilon =0.7$.
It is observed that a
Mel'nikov's soliton, moving with a velocity 
$\protect\upsilon _{\mathrm{sol}}=0.96$, is formed. The middle and 
bottom panels show snapshots of the
densities of the dark soliton in the $u$-component, and 
the bright soliton in the $\mathrm{v}$-component, for 
$t=170$ and $t=250$, respectively. Solid
(blue) and dashed (red) lines represent the numerical and analytical
results, the agreement between which is excellent.}
\label{fig4}
\end{figure}

On the other hand, in the intermediate velocity regime, 
$0.61<\upsilon <0.95$
[see the region between dashed (green)
lines in Fig.~\ref{fig2}], the $\mathrm{v}$-component forms a relatively
large peak, which is practically robust and does not decay in time.
Additionally, a dip of a constant density is formed in the large (${u}$)
component, which propagates with the same velocity as the peak in the the $%
\mathrm{v}$-component, i.e., the two components build a traveling coherent
structure. An example, shown in the top panel of Fig.~\ref{fig4} for 
$\upsilon =0.7$, 
reveals this in more detail: after a relatively
short transient period ($t\simeq 100$) needed for the $\mathrm{v}$-component
to reach the flat segment of the $u$-component, a small-amplitude
dark-bright soliton is formed.
Notice that the strong emission of sound waves --which is observed chiefly
in the $u$-component-- is solely due to the penetration process (as in the
case of Fig.~\ref{fig3}): the sound waves get detached from the subsonic
dark-bright soliton shortly after its generation, and the soliton propagates
undistorted thereafter over the spatial region occupied by the $u$-component.

To get a deeper insight into the solitary-wave formation, we apply an
asymptotic multiscale expansion method to Eqs.~(\ref{oldeq1}) 
and (\ref{oldeq2}). In particular, considering only the flat region 
of the $u$-component (where the soliton formation is observed), we neglect the
trapping potentials $V_{1,2}$ in Eqs.~(\ref{oldeq1})-(\ref{oldeq2}) (recall
that $V_{2}$ is switched off in the region where $V_{1}=0$ and the $u$%
-component is practically flat), and assume that chemical potential $\mu _{2}
$ is small. We thus introduce a formal small parameter $\varepsilon $ which
measures the smallness of the chemical potential of the $\mathrm{v}$%
-component, and substitute $\mu _{2}\rightarrow \varepsilon \mu _{2}$ in
Eqs.~(\ref{oldeq1})-(\ref{oldeq2}). Our main goal is to reduce the original
system of Eqs.~(\ref{oldeq1})-(\ref{oldeq2}) to a simpler one, namely the
completely integrable Mel'nikov system \cite{Mel}, which will provide an
analytical description of the vector soliton observed in Fig.~\ref{fig4}.

\begin{figure}[tbp]
\centering
\includegraphics[scale=0.37]{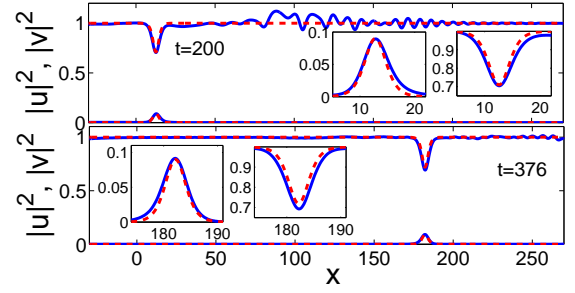}
\caption{(Color online) The same as in the middle and bottom panels of Fig.~%
\protect\ref{fig4}, but for a larger size of trap $V_{1}$, namely $%
L_{x_{1}}=123~\mathrm{\protect\mu }$m (or $L_{x_{1}}=490$ in dimensionless
units). In this case, as shown in the snapshots at $t=200$ (the top panel)
and $t=376$ (the bottom panel), the evolution time of the Mel'nikov soliton
is significantly longer. }
\label{fig5}
\end{figure}

The analysis starts by introducing the ansatz,
\begin{eqnarray}
u &=&\sqrt{\rho }\mathrm{exp}(i\varphi ),\quad \mathrm{v}=q\exp (i\theta ),
\label{eq11} \\
\theta  &=&\sqrt{\frac{\mu _{1}}{2}}x-\left[ \mu _{1}\left( \frac{1}{2}%
+\beta \right) +\varepsilon \Omega \right] t,  \label{eq12}
\end{eqnarray}
where real functions $\rho (x,t)$ and $\varphi (x,t)$ denote the amplitude
and phase of the $u$-component, while $q(x,t)$ and $\theta (x,t)$ represent
a (complex) amplitude and phase of the $\mathrm{v}$-component, with the real
parameter $\Omega $ setting the frequency of the $\mathrm{v}$-component. As we
will see below, $\Omega $ is related to the amplitude of the dark soliton in
the $u$-component.

Substituting Eqs.~(\ref{eq11})-(\ref{eq12}) into Eqs.~(\ref{oldeq1})-(\ref%
{oldeq2}), we obtain the following system of equations for fields $\rho $, $%
\varphi $ and $q$:
\begin{equation}
\varphi _{t}+\rho +\beta |q|^{2}-\mu _{1}+\varphi _{x}^{2}-\rho ^{-1/2}(\rho
^{1/2})_{xx}=0,  \label{redark}
\end{equation}%
\begin{equation}
\frac{1}{2}\rho _{t}+(\rho \varphi _{x})_{x}=0,  \label{imdark}
\end{equation}%
\begin{eqnarray}
i\left( q_{t}+\sqrt{2\mu _{1}}q_{x}\right) +q_{xx} &-&[|q|^{2}-\beta (\mu
_{1}-\rho )]q  \notag \\
&+&\varepsilon (\mu _{2}+\Omega )q=0.  \label{bright}
\end{eqnarray}%
Next, we define the slow variables:
\begin{equation}
T=\varepsilon ^{3/2}t,\quad X=\varepsilon ^{1/2}(x-c_{s}t),  \label{XT}
\end{equation}%
where $c_{s}$ is an unknown velocity, to be determined in a self-consistent
manner. Additionally, we express unknown fields $\rho $, $\varphi $ and $q$
as asymptotic series in $\varepsilon $:
\begin{eqnarray}
\rho  &=&\mu _{1}+\varepsilon \rho _{1}(X,T)+\varepsilon ^{2}\rho
_{2}(X,T)+\ldots ,  \label{ro} \\
\varphi  &=&\varepsilon ^{1/2}\varphi _{1}(X,T)+\varepsilon ^{3/2}\varphi
_{2}(X,T)+\ldots ,  \label{fi} \\
q &=&\varepsilon q_{1}(X,T)+\varepsilon ^{2}q_{2}(X,T)+\ldots   \label{Q}
\end{eqnarray}


Substituting expansions (\ref{ro})-(\ref{fi}) into Eqs.~(\ref{redark})-(\ref%
{imdark}), at the leading-order of approximation, i.e., at orders $\mathcal{O%
}(\varepsilon )$ and $\mathcal{O}(\varepsilon ^{3/2})$, compatibility
conditions of the resulting equations lead to the following results. First,
we determine the unknown velocity $c_{s}\equiv \sqrt{2}\tilde{c}_{s}$, 
which turns out to be the speed of sound of the $u$-component:
\begin{equation}
c_{s}=\pm \sqrt{2\mu _{1}},  \label{soundvel}
\end{equation}%
and derive an equation connecting unknown phase $\varphi _{1}$ with
amplitude $\rho _{1}$:
\begin{equation}
\varphi _{X}=c_{s}^{-1}\rho _{1}.  \label{fix}
\end{equation}%
Proceeding to the next orders, i.e., $\mathcal{O}(\varepsilon ^{2})$ and $%
\mathcal{O}(\varepsilon ^{5/2})$, the compatibility condition for the
respective equations ensuing from Eqs.~(\ref{redark})-(\ref{imdark}) yields
the following nonlinear equation:
\begin{equation}
2\rho _{1T}+\frac{6}{c_{s}}\rho _{1}\rho _{1X}-\frac{c_{s}}{2}\rho
_{1XXX}+c_{s}\beta (|q_{1}|^{2})_{X}=0.  \label{m1}
\end{equation}%
At the order $\mathcal{O}(\varepsilon ^{2})$, Eq.~(\ref{bright}) is reduced
to the following equation:
\begin{equation}
q_{1XX}-\beta \rho _{1}q_{1}+(\mu _{2}+\Omega )q_{1}=0.  \label{m2}
\end{equation}
\begin{figure}[tbp]
\centering
\includegraphics[scale=0.37]{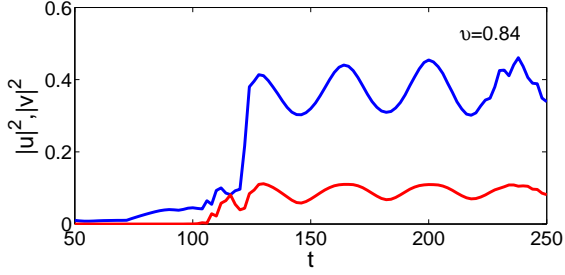}
\caption{(Color online) The evolution of the largest density dip in the $u$%
-component [the upper (blue) line] and peak density of the $\mathrm{v}$%
-component [the lower (red) line], for the trap's velocity $\protect\upsilon %
=0.84$.}
\label{fig6}
\end{figure}
The system of Eqs.~(\ref{m1})-(\ref{m2}) represents the Korteweg-de Vries
(KdV) equation with a self-consistent source, which is determined by a
stationary Schr\"{o}dinger equation, is the Mel'nikov system \cite{Mel},
which is \textit{completely integrable} and possesses exact soliton
solutions of the form:
\begin{eqnarray}
\rho _{1} &=&-\frac{2\delta }{\beta }\mathrm{sech}^{2}\eta ,  \label{rho1} \\
q_{1} &=&b\exp \left( -i\omega _{0}T\right) \mathrm{sech}\eta ,  \label{q1}
\end{eqnarray}%
where $\eta =\sqrt{\delta }(X+VT)$, with $V$ being the soliton velocity, and
$\omega _{0}$ is an arbitrary constant. 
Further, parameters $\delta \equiv \mu _{2}+\Omega $ and $b$
(which set the amplitudes of $\rho _{1}$ and $q_{1}$, respectively), are
related to the soliton velocity via the equation:
\begin{equation}
b^{2}=\frac{4\delta }{\beta ^{2}c_{s}}\left( V-\delta c_{s}\right) .
\label{b_gamma}
\end{equation}%
Obviously, the velocity of the Mel'nikov's soliton is bounded from below,
\textit{viz}., $V\geq V_{\mathrm{cr}}\equiv \delta c_{s}$, which follows
from the condition that $b^{2}$ is real.

Using the above results, we can now write down an approximate [valid up to
order $\mathcal{O}(\varepsilon )$] soliton solution of Eqs.~(\ref{oldeq1})-(%
\ref{oldeq2}). This solution is a vectorial soliton, composed of a dark
soliton in the $u$-component, coupled to a bright soliton in the $\mathrm{v}$%
-component. In terms of the original variables $x$ and $t$, it is expressed
as follows:
\begin{eqnarray}
\!\!\!\!\!
u(x,t) &\approx &\left( \mu _{1}-\frac{\varepsilon \delta }{\beta }\mathrm{sech}%
^{2}\eta \right) \exp \left( i\sqrt{\varepsilon \delta }c_{s}^{-1}\tanh \eta
\right) ,  
\label{ad} \\
\!\!\!\!\!
\mathrm{v}(x,t) &\approx &\varepsilon b\mathrm{sech}\eta \exp (i\theta ),
\label{ab}
\end{eqnarray}%
where $\eta \equiv \sqrt{\varepsilon \delta }(x-\upsilon _{\mathrm{sol}}t)$,
with $\upsilon _{\mathrm{sol}}\equiv c_{s}-\varepsilon V$ being the velocity
of the soliton. Note that this solution is characterized by two independent
parameters, which set the dark- and bright-soliton's amplitudes, $%
\varepsilon \delta $ and $\varepsilon b$. These parameters determine the
soliton's velocity too, via Eq.~(\ref{b_gamma}).
\begin{figure}[tbp]
\centering
\includegraphics[scale=0.37]{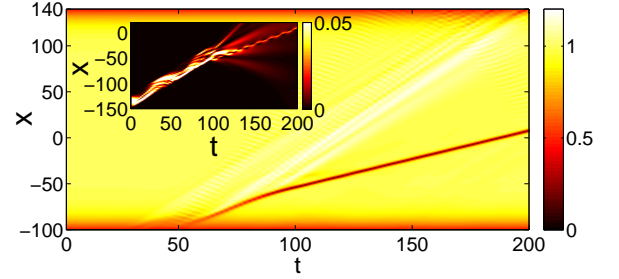}
\caption{(Color online) The same as in the top panel of Fig.~\protect\ref%
{fig4}, but for the trap's velocity $\protect\upsilon =1$. In this case,
although the $\mathrm{v}$-component is quickly dispersed (and dark-bright
solitons cannot be formed), it creates a robust dark soliton in the $u$%
-component. }
\label{fig7}
\end{figure}

This analytical result is compared to results of the numerical simulations
in the middle and bottom panels of Fig.~\ref{fig4}: upon numerically
determining the soliton's amplitudes, and finding its velocity by means of
Eq.~(\ref{b_gamma}), we can analytically determine the soliton density
profiles [dashed (red) lines] and compare them to the ones found by the
direct numerical integration of Eqs.~(\ref{oldeq1})-(\ref{oldeq2}) [solid
(blue) lines]. In this way, we find that the dark and bright soliton's
amplitudes are $\varepsilon \delta =0.27$ and $(\varepsilon b)^{2}=0.08$,
respectively, and the numerically found soliton velocity is $0.88$. The
latter value is in excellent agreement with the analytical prediction, $%
\upsilon _{\mathrm{sol}}=0.89$. This is also directly observed in Fig.~\ref%
{fig4}, where the analytical predictions [dashed (red) lines] are compared
to results of the simulations [solid (blue) lines], see the snapshots
corresponding to $t=170$ (middle panel) and $t=250$ (bottom panel). A
similar excellent agreement is observed in Fig.~\ref{fig5}, but for a longer
propagation time (for trap $V_{1}$ of a larger size, $L_{x_{1}}=123~\mathrm{%
\mu }$m, or $L_{x_{1}}=490$ in the dimensionless notation). We note in
passing that we have also checked that the Mel'nikov dark-bright soliton
remains robust after its reflection from the trap boundaries and its
interaction with the sound waves.

Coming back to Fig.~\ref{fig2}, it is observed that, close to the right edge
of the intermediate velocity regime, \textit{viz}., at $0.8\lesssim \upsilon
<0.95$, the solitons undergo a transient behavior: their peak densities
perform breathing oscillations, as shown in Fig.~\ref{fig6} for $\upsilon
=0.84$. Clearly, this type of the behavior cannot be captured by the
Mel'nikov model (which does not give rise to breathing modes). After this
transient region, in the regime of a large trap velocity, $\upsilon >0.95$,
the $\mathrm{v}$-component does not stay localized and is quickly dispersed;
this can also be observed in Fig.~\ref{fig2}. Thus, in this case, the
dark-bright soliton of the  Mel'nikov type is not formed. However, shortly
after penetrating the $u$-component, the $\mathrm{v}$-component creates a
dark soliton there, which propagates robustly all over the spatial region
occupied by the $u$-component, cf. Fig.~\ref{fig7} for $\upsilon =1$.

To summarize, in this case of the $v$-component trap was switched
off, we have identified three basic regimes. For small initial
trap's speeds, the $u$ and $v$ components are too slow for combining
into one of the higher-speed Mel'nikov-type dark-bright solitons. For
the trap's speed which is too large, the intrusion of the $v$
component into $u$ produces a single dark soliton, but subsequently
the $v$ component disperses fast, without creation of other coherent
structures. In the intermediate regime, the two components can
``lock'' into a dark-bright solitary wave, which can be mapped into
the solitons of the Mel'nikov model.

\subsection{Evolution of 
the small component in the presence of the trap.}

\begin{figure}[tbp]
\centering
\includegraphics[scale=0.37]{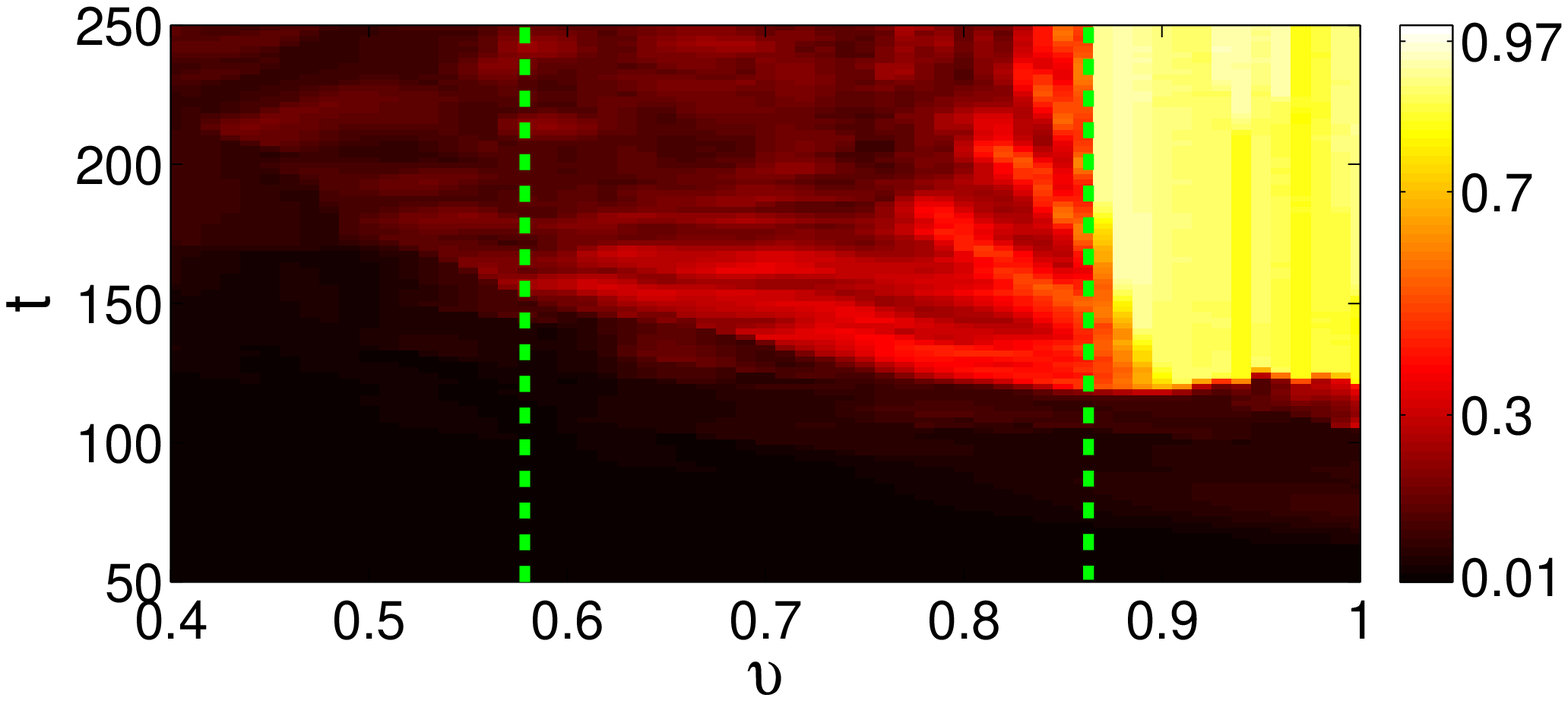} %
\includegraphics[scale=0.37]{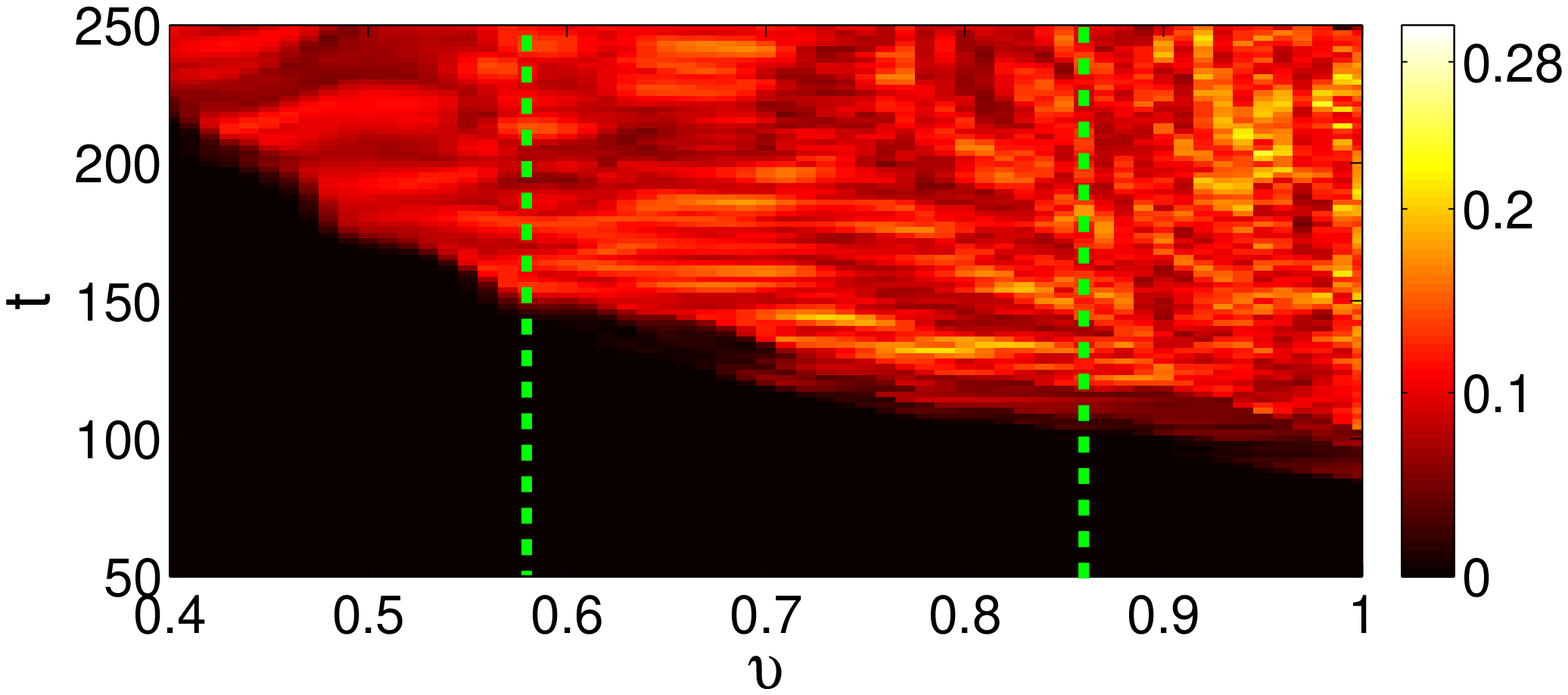}
\caption{(Color online) The same as in Fig.~\protect\ref{fig2}, but in the
case when the $\mathrm{v}$-component is subject to the action of the trap.
In this case too, three different regimes can be identified: a) 
$0<\protect \upsilon <0.58$
with no localized excitations, b) $0.58<\upsilon <0.87$, 
where unstable dark-bright
solitons are observed, and c) $\protect\upsilon >0.87$,
in which case we observe the nucleation of multiple dark solitons 
in the $u$-component.}
\label{fig8}
\end{figure}

We now consider the case when trap $V_{2}$ is not switched-off and, thus,
the $\mathrm{v}$-component evolves in the presence of its trap. The
pertinent contour plots depicting the largest dip in the ${u}$-component
and the peak density of the $\mathrm{v}$-component are shown in the top and
bottom panels of Fig.~\ref{fig8}, respectively. Similarly to the case
without the trap, three different regimes can be identified. In particular,
for low trap velocities $\upsilon <0.58$, 
the dynamics of the
two components is similar to that observed in Fig.~\ref{fig3}. Again, in
this case, long-lived localized structures are not formed in either
component, and strong emission of radiation is observed.

\begin{figure}[t]
\centering
\includegraphics[scale=0.37]{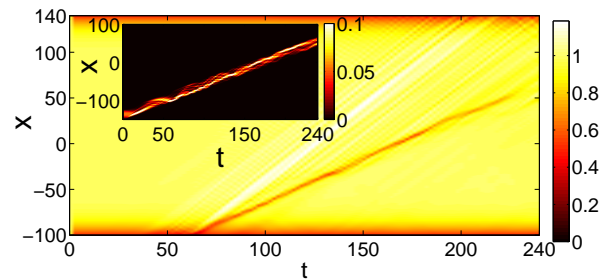}
\caption{(Color online) The same as in the top panel of Fig.~\protect\ref%
{fig4}, but for the trap's velocity $\protect\upsilon =0.8$.
In this case, a dark-bright soliton of the Mel'nikov type is formed, but it
decays due to the fact that the soliton's velocity is larger than the speed
of the trap (see text). }
\label{fig9}
\end{figure}

For the trap velocity in the intermediate region of 
$0.58<\upsilon <0.87$ 
[see the region in between the two dashed
(green) lines in Fig.~\ref{fig8}], we find that dark-bright solitons of the
the Mel'nikov type are again formed (as in the previous case where the trap $%
v_{2}$ was switched off), but they eventually decay. A typical example of
this behavior is shown in Fig.~\ref{fig9}, for $\upsilon =0.8$. As is
observed in the inset of the figure, after the $\mathrm{v}$-component has
intruded into the $u$-field, it splits into fragments, following its in-trap
dynamics. The largest one among these fragments couples to a dark soliton in
the $u$-component, and the resulting structure can be categorized as a
Mel'nikov soliton: for the parameters used in Fig.~\ref{fig9}, the
numerically found average dark and bright soliton's amplitudes are $%
\varepsilon \delta \approx 0.26$ and $(\varepsilon b)^{2}\approx 0.08$,
which correspond to the soliton's velocity [found via Eq.~(\ref{b_gamma})] 
$\upsilon_{\mathrm{sol}}\approx 1.1$. This is in good agreement with the numerically
found mean soliton's velocity, which is $\approx 1$, a fact that indicates
that this structure is indeed proximal to a Mel'nikov soliton. However, in
the case of Fig.~\ref{fig9}, the $\mathrm{v}$-component travels with speed $%
\upsilon =0.8$, while the soliton propagates with a larger velocity.
Therefore, the confined bright-soliton component inevitably collides
with -- and is reflected by -- the trap boundaries. This results in
complex motion of the dark-bright soliton, which cannot be sustained
for long times and eventually decays (at $t\sim 200$). In fact, it
is the trap itself in this case which is detrimental to the
existence of the Mel'nikov soliton. While the latter has its
intrinsic velocity upon the formation, the trap bears its own
distinct velocity, and, barring a non-generic scenario when these
two velocities coincide, we cannot expect the solitary wave to
persist, due to the unavoidable collision with either the ``front"
or the ``back" of the trap. This is the key distinguishing feature
between the intermediate speed cases in the presence and absence of
the trap.

\begin{figure}[tbp]
\includegraphics[scale=0.345]{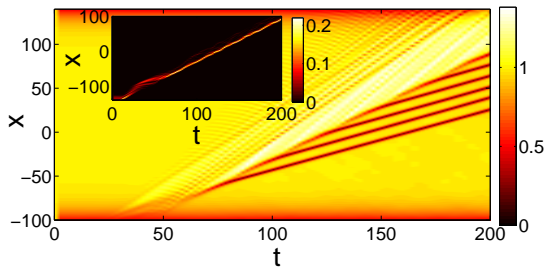}
\caption{(Color online) The top panel is the same as in Fig.~\protect\ref%
{fig4}, but for the trap's velocity $\protect\upsilon =1.2$.
In this case, the superfluidity of the $u$-component breaks down, and multiple
dark solitons are formed. }
\label{fig10}
\end{figure}

When the trap's velocity exceeds the value $\upsilon =0.87$,
we observe a constant robust large dip, in the ${u}$-component, while 
the $\mathrm{v}$-component features a peak, oscillating around 
a relatively large value. As we show below, this is a critical 
velocity above which multiple dark solitons are generated in the 
${u}$-component, indicating breakdown of the superfluidity.
As shown in the top panel of Fig.~\ref{fig8}, a persistent, almost
zero-density region arises in the ${u}$-component. In this case, the
nucleation of multiple dark solitons in the $u$-component occurs. A typical
example is shown in the top panel of Fig.~\ref{fig10} (for $\upsilon =1.2$):
shortly after the $\mathrm{v}$-component has intruded into
the flat part of the $u$-component, dark solitons are repeatedly emitted.
This behavior can be understood as the breakdown of the superfluidity in the
$u$-component, caused by the supersonic motion of the $\mathrm{v}$%
-component: in fact, as shown in the bottom panel of
Fig.~\ref{fig10}, at early stages of the evolution, the
$\mathrm{v}$-component actually becomes a sharply localized
``obstacle", which moves with a velocity exceeding the critical one
(which is, roughly, the speed of sound $c_{s}$ of the
$u$-component). Accordingly, the dynamics follows the scenario
revealed in previous theoretical \cite{hakim,thsf} and experimental
\cite{engelsath} works concerning the breakdown of the superfluidity
in quasi-1D BECs. In all these works, continuous emission of dark
solitons was reported. The mechanism of the breakdown was elucidated
in the pioneering work of~Hakim \cite{hakim}. In particular, up to
the critical point, the
moving defect (in our case, the effective one, represented by the $v$%
-component) can sustain a pair of localized states co-traveling with it (a
saddle and a center). Past the critical point, a saddle-center bifurcation
arises, and such states can no longer be sustained. To alleviate the local
super-criticality (in terms of the speed), the defect nucleates a
(sub-sonic) dark solitary wave. Yet, once the wave has traveled sufficiently
far downstream, it no longer renders the vicinity of the defect
sub-critical, and a new dark soliton is emitted. The repetition of this
process leads to the emergence of a dark-soliton train.

\begin{figure}[tbp]
\includegraphics[scale=0.345]{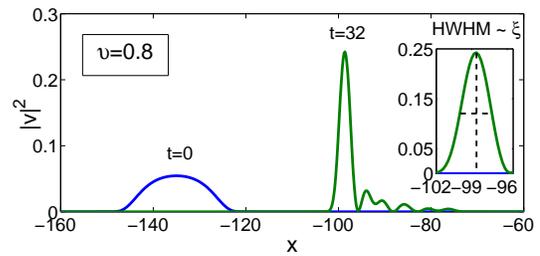}
\caption{(Color online) The plot shows the density of the 
$\mathrm{v}$-component for the trap's velocity 
$\protect\upsilon=0.8$,
in two instants: at $t=0$ (the blue line) and at
$t=32$ (the green line), when it attains the maximum peak density.
The inset shows the density profile of the larger part of the
$\mathrm{v}$-component at $t=32$; its half-width at half-maximum
(HWHM) equals to $1.3$, being on the order of the healing length,
$\protect\xi $. }
\label{fig11}
\end{figure}

It is relevant to quantitatively justify the above arguments, upon
estimating the critical velocity needed for the breakdown of the
superfluidity. This can be done as follows. First, we observe, in
the example depicted in Fig.~\ref{fig11} for $\upsilon=0.8$,
that the $\mathrm{v}$-component has initially
(at $t=0$) the form of an extended wave packet (shaped by trap
acting on it); however, after intruding into the $u$-field, 
it splits into fragments. 
The largest among these fragments has the peak density approximately
four times as large as the initial one, and becomes sharply
localized, with the half-width at half-maximum of the order of
healing length $\xi $ (which is equal to $1$ in our dimensionless
notation). Based on this observation, it is possible to consider the
$\mathrm{v}$-component as a delta-like obstacle of some effective
strength $\mathrm{v}_{\ast }$. To calculate it, we fit the
$\mathrm{v}$-component by a normalized Gaussian
function with an equal amplitude at $t\sim 32$ (when the 
$\mathrm{v}$-component attains its maximum peak density). 
Then, we use the approximate expression for the critical 
velocity of Ref.~\cite{hakim},
which reads:
\begin{equation}
\mathrm{v}_{\ast }=4(1-\upsilon _{\mathrm{cr}}^{2}/2)\frac{[\sqrt{%
1+4\upsilon _{\mathrm{cr}}^{2}}-(1+\upsilon _{\mathrm{cr}}^{2})]^{1/2}}{%
2\upsilon _{\mathrm{cr}}^{2}-1+\sqrt{1+4\upsilon _{\mathrm{cr}}^{2}}}.
\label{hak}
\end{equation}%
%

\begin{figure}[tbp]
\includegraphics[scale=0.345]{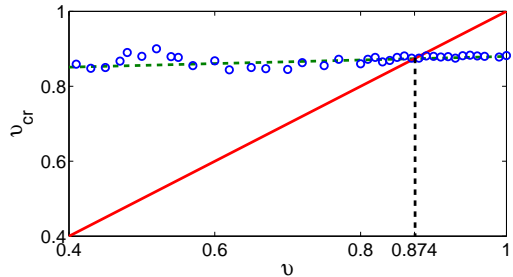}
\caption{(Color online) The critical velocity, $\protect\upsilon _{\mathrm{cr%
}}$, versus the trap's velocity, $\protect\upsilon $. First we calculate the
value of $\protect\upsilon _{\mathrm{cr}}$ [depicted by (blue) circles] for
different trap's velocities, and then apply the straight-line fit [the
dashed (green) line]. Breakup of the superfluidity is expected when the
trap's velocity becomes larger than the critical velocity, i.e. at 
$\protect \upsilon \approx 0.874$,
as indicated by the vertical dashed
(black) line. }
\label{fig12}
\end{figure}

In Fig.~\ref{fig12}, we show the dependence of the critical velocity $%
\upsilon _{\mathrm{cr}}$, calculated as explained above, on the trap's
velocity $\upsilon $. By applying a straight-line fit to the values of $%
\upsilon _{\mathrm{cr}}$, we find that $\upsilon _{\mathrm{cr}}\approx
0.048\upsilon +0.832$. Thus, there is a very weak dependence of the
effective strength of amplitude of the steep moving obstacle, represented by
the $\mathrm{v}$-component, on the initial trap's speed. By comparing this
fit with the diagonal in the (positive quarter-) plane of $\upsilon _{%
\mathrm{cr}}$ vs. $\upsilon $, we can conclude what initial speeds of the
trap turn out to be supercritical. In particular, it is clear from Fig.~\ref%
{fig12} that, for $\upsilon >0.874$,
the trap exceeds the
critical velocity $\upsilon _{\mathrm{cr}}$, hence the superfluidity is
expected to break down. Note that the numerically obtained velocity, beyond
which the superfluidity breaks down indeed and dark solitons are emitted was
found to be $\upsilon =0.87$
[see the rightmost dashed (green)
lines in Fig.~\ref{fig8}], which is in excellent agreement with the
theoretical approximation based on Eq.~(\ref{hak}).

In summary, three regimes are also identified in the case when the $\mathrm{v%
}$-component is subject to the action of the potential, although their
dynamics is different from the case when the trap was absent, except for the
low-speed case. In the intermediate-speed regime, solitary waves of the
Mel'nikov type form but are unstable due to the velocity mismatch with the
moving trap. In the large-speed regime, the superfluidity was shown to break
down at the critical point very close to the related analytical prediction,
which leads, as in the case of~Ref. \cite{hakim}, to the emission of a train
of dark solitons, due to the effective potential exerted by the second ($%
\mathrm{v}$) component on the first ($u$) one.

\section{Conclusions.}

In this work, we have studied the dynamics of two quasi-1D
counter-propagating immiscible superfluids. Our setup is based on the BEC
mixture composed by two different spin states of the same atomic species,
with the number of atoms in one of the two components being much smaller
than in the other. 
In our simulations of the coupled GPEs, we have assumed that the
``large" component is at rest while the ``small" component starts
intruding into the larger one, moving at a constant velocity
$\upsilon $.

We have considered two different situations, namely, with the small
component propagating either in the absence or in the presence of the trap.
For sufficiently small trap's velocities $\upsilon$,
we have found that no localized structures emerge.

On the other hand, for intermediate velocities in the interval of 
$0.61\tilde{c}_{s}<\upsilon <0.95\tilde{c}_{s}$, if the trap acting on the small
component is switched off, we have found that a robust dark-bright soliton
is formed. This structure is composed of a bright (dark) soliton in the
small (large) component, and can be very well approximated analytically by
means of the multiscale asymptotic expansion method, which reduces the
original GPE\ system to the completely integrable Mel'nikov system. This
system has exact soliton solutions that can be used to construct the
approximate dark-bright soliton solutions of the GPE system. We have also
found that, if the small component is subject to the action of the trapping
potential, then the Mel'nikov-type dark-bright soliton forms in a similar
velocity regime, $0.58\tilde{c}_{s}<\upsilon <0.87\tilde{c}_{s}$, but it
cannot be sustained due to the mismatch between its own velocity and the
speed of the moving trap. 

If the small component propagates in the absence of the trap, then, for
large  initial trap velocities, $\upsilon >0.95\tilde{c}_{s}$, it disperses
and no dark-bright soliton is formed; nevertheless, a dark soliton is always
created in the large component. On the other hand, if the small component
propagates in the presence of its trap, which moves with speed $\upsilon
>0.87\tilde{c}_{s}$, we have found that it gets deformed into a sharply
localized object of the width on the order of the healing length, which
propagates with a velocity larger than the critical velocity of the large
component. Approximating the small component by an effective delta-like
obstacle acting on the large component, we have found that, in the case of
such large trap velocities, the dynamics of the system follows the scenario
known for the quasi-1D superfluid flow past an obstacle. In this way, the
superfluidity of the large component breaks down, and nucleation of a train
of dark solitons occurs.

Our results suggest that such counterflow settings may be quite relevant to
the observation of fundamental phenomena, such as the formation of solitons
and/or the identifying the critical velocity in superfluid flows. It would
be interesting to extend the considerations to higher-dimensional settings
and, also, to larger number of components, as, e.g., in the case of $F=1$ or
$F=2$ spinor BECs~\cite{ueda_review}. In that case, vortex-bright solitary
waves~\cite{kody} and their dipole-more generalizations~\cite{martina} are
expected to arise, as well as their spinor counterparts. The generalizations
with a larger number of components and in higher dimensions is likely to
produce novel waveforms. Another relevant extension concerns the role of 
$g_{12}$ (the cross-component interaction strength) and the approach to the
miscibility-immiscibility threshold. In that context, exploring the
conditions under which additional waveforms (such as dark-dark solitons 
of~\cite{darkdark1,darkdark2} and their multi-dimensional counterparts) can arise would
also be a theme of interest. These topics are currently under investigation
and will be reported elsewhere.

\textbf{Acknowledgments.} Constructive discussions with P. Engels are kindly 
acknowledged. V.A. appreciates financial support from A.~G.
Leventis Foundation. The work of D.J.F. was partially supported by the
Special Account for Research Grants of the University of Athens. P.G.K.
gratefully acknowledges support from the Alexander von Humboldt Foundation
and from the NSF under grant DMS-0806762. The work of P.G.K. and B.A.M. was
supported, in a part, by the Binational (US-Israel) Science Foundation under
grant No. 2010239.


\end{document}